\documentclass[secnumarabic, graphics,floatfix, nofootinbib,tightenlines,nobibnotes, aps, prl, 12pt]{revtex4-1}
\usepackage{graphicx}
\usepackage[english]{babel}
\usepackage[utf8]{inputenc}
\usepackage{amsmath,amssymb}
\usepackage{amsfonts}
\usepackage{multirow}
\usepackage{pstricks}
\usepackage{float}
\usepackage{subfigure}
\usepackage{color}
\usepackage{epsfig}
\usepackage{pst-tools}

\begin{document}

\title{A quasiparticle equation of state with a phenomenological critical point}
\author{Hong-Hao Ma$^1$}
\author{Wei-Liang Qian$^{2,1}$}

\affiliation{$^1$Faculdade de Engenharia de Guaratinguet\'a, Universidade Estadual Paulista, 12516-410, Guaratinguet\'a, SP, Brazil}
\affiliation{$^2$Escola de Engenharia de Lorena, Universidade de S\~ao Paulo, 12602-810, Lorena, SP, Brazil}

%\date{\today}
\date{Dec. 03, 2017}

\begin{abstract}
We propose a hybrid parameterization of a quasiparticle equation of state, where a critical point is implemented phenomenologically.
In this approach, a quasiparticle model with finite chemical potential is used to describe the quark-gluon plasma phase by fitting to the lattice quantum chromodynamics data at high temperature.
On the other hand, the hadronic resonance gas model with excluded volume correction is employed for the hadronic phase.
An interpolation scheme is implemented so that the phase transition is a smooth crossover when the chemical potential is smaller than a critical value, or otherwise approximately of first order according to Ehrenfest's classification.
Also, the thermodynamic consistency is guaranteed for the equation of state related to both the quasiparticle model and the implementation of the critical point.

\pacs{12.38.Bx, 12.38.Aw, 11.15.Bt}

\end{abstract}
\maketitle

\section{Introduction}

Quasiparticle model provides a phenomenological approach to the thermodynamic properties of quark-gluon plasma (QGP) obtained by lattice quantum chromodynamics (QCD).
It is part of the efforts to identify the appropriate number of degrees of freedom of the system for the region where nonperturbative effects become dominant.
In this approach, the strongly interacting matter is interpreted as consisting of non-interacting quanta carrying the same quantum numbers of quarks and gluons.
As inspired by its counterparts in other fields of physics, the strong interactions among the constituents of the system are incorporated through the temperature dependent effective masses.
The quasiparticle model was first proposed by Peshier $et~al.$~\cite{Peshier:1994zf}.
It is reformulated by Gorenstein and Yang~\cite{Gorenstein:1995vm} to achieve the thermodynamical consistency, via the introduction of a temperature dependent bag constant.
The latter is determined by canceling the additional term emerging in the thermodynamic constraint relation owing to the temperature dependent mass.
Thereafter, many alternative thermodynamically consistent approaches have been proposed~\cite{Biro:2001ug,Bannur:2006hp,Bannur:2006js,Bannur:2006ww,YIN:2008ME,YIN:2008MH,Oliva:2013zda}.
In the model proposed by Bannur~\cite{Bannur:2006hp,Bannur:2006js,Bannur:2006ww}, the form of the internal energy as well as the particle number are taken to preserve their respective forms in statistical mechanics.
The pressure, as well as other thermodynamic quantities, are then obtained by the standard procedure of statistical mechanics, which are shown to be consistent with the thermodynamic relation.
In this model, temperature dependent bag constant is not introduced as an {\it a priori} assumption.
Moreover, if one chooses a particular value for the constant of integration in the thermodynamic relation, Gorenstein and Yang's formalism is manifestly restored~\cite{Bannur:2006hp}.

Lattice QCD studies~\cite{lattice-01,lattice-02} showed that for vanishing baryon density and large strange quark mass the transition is a smooth crossover.
At non-vanishing chemical potential, on the other hand, a variety of model calculations~\cite{Halasz:1998qr,Berges:1998rc,Stephanov:1998dy,Schwarz:1999dj,Fodor:2004nz} indicated the existence of a first order phase transition.
These results imply that the phase diagram is probably featured by a critical point where the line of first-order phase transitions terminates, and the transition is expected to be of second order at this point.
As a matter of fact, the existence and properties of the critical point is a long-standing intriguing topic.

In order to study the effect of the equation of state (EoS) in heavy ion collisions, Huovinen and Petreczky proposed a parameterization~\cite{Huovinen:2009yb} which combines the hadron resonance gas (HRG) model at low temperature with the lattice QCD data at high temperature~\cite{Bazavov:2009zn}.
In their approach, an inverse polynomial fit is utilized for the lattice data, and it is matched to the HRG model at the joining temperature $T_0$ by requiring that the trace anomaly, as well as its first and second derivatives, are continuous.
Thereafter, their parametrization was widely used in hydrodynamical model calculations.
However, the EoS mentioned above only applies to zero baryon chemical potential.
As a result, it does not provide the possibility to investigate the properties of finite baryon density, and in particular, those regarding the critical point where the transition evolves from a smooth crossover to a first order phase transition.
Also, various Lattice QCD groups have improved their calculations, and new EoS data were published in the past few years~\cite{Bazavov:2014pvz,Borsanyi:2012cr,Borsanyi:2013bia,Bellwied:2015rza,Borsanyi:2016ksw}.

These concerns motivated the present study of a hybrid EoS to take into account these aspects on a phenomenological level.
In our approach, the QGP phase is connected to the hadronic phase with the introduction of a phenomenological critical point.
We employ a quasiparticle model with finite chemical potential proposed by Bannur~\cite{Bannur:2006hp,Bannur:2006js,Bannur:2006ww,Bannur:2007tk} to describe the QGP phase.
The parameter of the model is adjusted to reproduce the recent Lattice QCD results of stout action~\cite{Borsanyi:2012cr,Borsanyi:2013bia}.
At low temperature, an HRG model with excluded volume correction~\cite{Rischke:1991ke,Hama:2004rr} is utilized for the description of the hadronic phase.
Additionally, a critical point is implemented phenomenologically at finite baryon chemical potential.
The latter is achieved by adopting the interpolation scheme proposed by Hama $et~al.$~\cite{Hama:2005dz}.

The present work is organized as follows.
In section II, we briefly review the quasiparticle model employed in this work and discuss the model parameterization.
The HRG model is presented in section III.
The interpolation scheme for the phenomenological critical point is studied in section IV.
We present the numerical results in section V together with some discussions.
Concluding remarks are given in the last section.

\section{Quasiparticle model for 2+1 flavor QGP}

To reproduce the lattice QCD data at high temperature~\cite{Bazavov:2014pvz}, in this work we employ the quasiparticle model proposed in~\cite{Bannur:2007tk}.
An important aspect of the approach is that it does not introduce a temperature dependent bag constant which satisfies a restrictive condition~\cite{Gorenstein:1995vm,Bannur:2006hp}.
The approach keeps the form of energy and particle number the same as they are formulated as grand ensemble averages in statistical mechanics as follows,

\begin{eqnarray}
E \equiv \langle E_i \rangle= \frac{\sum\limits_i E_i \exp (-\alpha N_i-\beta E_i)}{\sum\limits_i \exp (-\alpha N_i-\beta E_i)} , \nonumber\\
N \equiv \langle N_i \rangle= \frac{\sum\limits_i N_i \exp (-\alpha N_i-\beta E_i)}{\sum\limits_i \exp (-\alpha N_i-\beta E_i)} .
\end{eqnarray}
The above expression can be rewritten in terms of the grand partition function,
\begin{eqnarray}
Q_{G}=\sum\limits_i \exp (-\alpha N_i-\beta E_i) ,
\end{eqnarray}
where each microstate is labelled by $i$, which has total particle number $N_i$ and total energy $E_i$. To be specific,
\begin{eqnarray}
\langle E_i \rangle &=& -\frac{\partial }{\partial \beta}\left\{\ln \sum\limits_i \exp (-\alpha N_i-\beta E_i)\right\}-\beta \frac{\partial m}{\partial \beta}\langle \frac{\partial E_i}{\partial m}\rangle , \nonumber\\
\langle N_i \rangle &=& -\frac{\partial }{\partial \alpha}\left\{\ln \sum\limits_i \exp (-\alpha N_i-\beta E_i)\right\} .
\end{eqnarray}
Here we note there is an extra term involving $\left(\frac{\partial m}{\partial \beta}\right)$ in the expression for the ensemble average of energy owing to the temperature denpendence of quasiparticle mass.

Following the standard procedure of statistical mechanics \cite{Pathria1996}, other thermodynamic quantities are subsequently identified by matching the total derivative of $q=\ln Q_G$ to the first law of thermodynamics.
To be specific, one has
\begin{eqnarray}
dq=-\langle N_i \rangle d\alpha - \langle E_i \rangle d\beta - \beta \langle \frac{\partial E_i}{\partial V} \rangle dV - \beta \frac{\partial m}{\partial \beta}\langle \frac{\partial E_i}{\partial m}\rangle d\beta .
\end{eqnarray}
By comparing the above expression with the first law of thermodynamics, it can be inferred that
\begin{eqnarray}
&&\beta = \frac{1}{k_{B}T} , \nonumber \\
&&\alpha = -\frac{\mu}{k_{B}T} , \nonumber \\
&&q+\alpha N +\beta E + \int d\beta\beta\frac{\partial m}{\partial \beta}\langle \frac{\partial E_i}{\partial m}\rangle = \frac{S}{k_{B}} .
\end{eqnarray}
Subsequently, one finds the expression for pressure,
\begin{eqnarray}
\frac{pV}{k_{B}T} = \frac{(E+pV-TS)-E+TS}{k_{B}T} = \frac{\mu N-E+TS}{k_{B}T} = q + \int d\beta\beta\frac{\partial m}{\partial \beta}\langle \frac{\partial E_i}{\partial m}\rangle \label{eqP} .
\end{eqnarray}
It is readily to verify \cite{Bannur:2006hp} that Eq.(\ref{eqP}) is in consistency with the thermodynamical relation
\begin{eqnarray}
\epsilon \equiv \frac{E}{V} = T\frac{\partial p}{\partial T} - p .
\end{eqnarray}

\subsection{Parameterization for 2+1 flavor QGP at zero chemical potential}

The 2+1 flavor QGP consists of a system of non-interacting quasiparticles carrying the quantum numbers of the gluons, the up, down as well as strange quarks.
The single particle energy of quasiparticles $\omega_k$ depend on thermal mass and momentum $k$.
Here we consider the on-shell dispersion relation
\begin{eqnarray}
\omega_k^2=k^2+m_{g,q}^2  ,
\end{eqnarray}
where the following prescription~\cite{Bannur:2007tk,Levai:1997yx} for the thermal masses of quasiparticles are adopted, $i.e.$,
\begin{eqnarray}
m_{g}^{2}=\frac{3}{2} \omega_{p}^{2}
\end{eqnarray}
for gluons and
\begin{eqnarray}
m_{q}^{2}=(m_{q0}+m_f)^2 + m_{f}^{2}
\end{eqnarray}
for quarks, where $q$ stands for $u$, $d$, or $s$ quark.
Here $m_{q0}$ stand for the current mass of the quarks.
We take $m_{s0}=0.150$ GeV for strange quark, and $m_{u0,d0}=m_{s0}$/28.15 $\approx$ 5.33 MeV for up and down quarks.
The plasmon frequency $\omega_p$ and the effective mass of soft massless quark $m_f$ are associated with the collective behavior of the system.
They can be obtained by analysing the poles of the relevant propagators using the hard thermal loop (HTL) approximation~\cite{Pisarski:1989cs,Frenkel:1989br},
\begin{eqnarray}
\omega_{p}^{2}=&&\frac{g^2 T^2}{18} (2N_c+n_f), \label{qcdmass1}\\
m_{f}^{2}=&&\frac{N_{c}^{2}-1}{2N_c} \frac{g^2 T^2}{8}, \label{qcdmass2}
\end{eqnarray}
where the number of colors $N_c=3$, the number of flavors $n_f=2+1=3$ and $g$ is the coupling constant to be specified below.
For the low temperature region, we adopt the parameterization of model II proposed in \cite{Bannur:2007tk} as follows
\begin{eqnarray}
\omega_{p}^{2}=&&a_{g}^{2} g^2 \frac{n_g}{T} + \sum_{q} a_{q}^{2} g^2 \frac{n_q}{T}, \label{vmbmass1}\\
m_{f}^{2}= &&b_{q}^{2} g^2 \frac{n_q}{T}, \label{vmbmass2}
\end{eqnarray}
where $n_g$ and $n_q$ are number densities of gluons and quarks.
Here the coefficients $a_g$, $a_q$ and $b_q$ are to be determined by demanding Eqs.(\ref{vmbmass1}-\ref{vmbmass2}) approach the perturbative results, Eqs.(\ref{qcdmass1}-\ref{qcdmass2}), as $T \rightarrow \infty$.

The principle of asymptotic freedom indicates that the effective coupling constant decreases as the momentum transfer increases.
In a thermal medium, the characteristic momentum transfer between quanta is of the order of the temperature.
Therefore, the coupling constant $g$ falls with increasing temperature, as obtained by the two-loop approximation~\cite{Caswell:1974gg,BeiglboCk:2006lfa},
\begin{eqnarray}
\alpha_s(T) \equiv \frac{g^2}{4 \pi}=\frac{6 \pi}{(33-2n_f)\ln (T/\Lambda_T)}(1-\frac{3(153-19n_f)}{(33-2n_f)^2}\frac{\ln (2\ln (T/\Lambda_T))}{\ln (T/\Lambda_T)}) \label{alphas0} .
\end{eqnarray}

The above system of coupled equations thus can be solved self-consistently for plasma frequency and number density, where the energy density and the number density of the 2+1 flavor QGP read
\begin{eqnarray}
\varepsilon=&& \varepsilon_{g} +\varepsilon_{u} +\varepsilon_{d} +\varepsilon_{s}, \\
n=&&n_{g} +n_{u}+n_{d} +n_{s}.
\end{eqnarray}
where
\begin{eqnarray}
\varepsilon_{i}=&& \frac{g_i}{2 \pi^2} \int_{0}^{\infty} dk \frac{(k^2+m_{i}^{2})k^2}{e^{(\sqrt {k^2+m_{i}^{2}}-\mu_i)/T}\pm 1}+(\mu_i \rightarrow -\mu_i)\equiv \varepsilon_{i}^{id}, \\
n_{i}=&&\frac{g_i}{2 \pi^2} \int_{0}^{\infty} dk \frac{k^2}{e^{(\sqrt {k^2+m_{i}^{2}}-\mu_i)/T}\pm 1}-(\mu_i \rightarrow -\mu_i)\equiv n_{i}^{id}.
\end{eqnarray}
where ``$-$" in the denominator applies to bosons and ``$+$" is for fermions, and $g_i$ is the degeneracy.
For the present case of zero chemical potential, $\mu_i=0$, the number density vanishes identically.
As discussed before, the pressure can be calculated by using the thermodynamic relation,
\begin{eqnarray}
\frac{p}{T}=&&\frac{p_0}{T_0}+ \int_{T_0}^{T} dT \frac{\varepsilon(T)}{T^{2}}.
\end{eqnarray}
where $p_0$ and $T_0$ are the pressure and temperature at some reference points.
Here we choose $T_0=0.175$ and ${p_0}/{T_{0}^{4}}=1.08$ respectively.

For the case of zero chemical potential, the temperature related scale parameter is taken to be $\Lambda_T=0.135$ GeV.
Owing to the $\ln(T/\Lambda_T)$ term in Eq.(\ref{alphas0}), the expression is not well defined when $T \le \Lambda_T$, thus an extrapolation is employed for the region $T \lesssim \Lambda_T$.
Numerical calculations show that the contributions from the HRG dominate in the region, and subsequently, the results are not sensitive to any particular choice of extrapolation scheme.

\subsection{Parameterization for 2+1 flavor QGP at finite chemical potential}

Following \cite{Bannur:2006js,Schneider:2003uz}, for finite chemical potential, the term $T/\Lambda_T$ in Eq.(\ref{alphas0}) can be replaced by
\begin{eqnarray}
\frac{T}{\Lambda_T}= \frac{T}{\Lambda_T} \sqrt{1+(1.91/2.91)^2 \frac{\mu^2}{T^2}} .
\end{eqnarray}
Also, the plasma frequencies are replaced by \cite{Peshier:1999ww}
\begin{eqnarray}
m_{f}^{2}=&& \frac{g^2 T^2}{18}n_f (1+\frac{\mu^{2}}{\pi^2 T^2}) \label{mfmu}.
\end{eqnarray}

The pressure can be determined via an integral from its value at zero chemical potential
\begin{eqnarray}
\Delta p =p(T,\mu)-p(T,0)= \int_{0}^{\mu} n_q d\mu
\end{eqnarray}
Here number density $n_q$ can be calculated by taking accont into the modified plasma frequencies as well as the chemical potential.
Other thermodynamic quantities are obtained according to the thermodynamic relations
\begin{eqnarray}
&&\Delta s=\partial \Delta p/ \partial T,\\
&&\Delta \varepsilon=T \Delta s- \Delta p + \mu_B n_B + \mu_S n_S.
\end{eqnarray}

In the present study, we consider strangeness neutrality condition.
Since the strangeness solely comes from strange quark, strangeness neutrality implies $\mu_s=0$.
Therefore, for light quarks, we take $\mu_u=\mu_d=\mu_B/3$.
One sees that Eq.(\ref{mfmu}) restores Eq.(\ref{qcdmass2}) at vanishing chemical potential.
However, in our present study, we employ Eq.(\ref{vmbmass2}) which only approaches Eq.(\ref{qcdmass2}) as $T\rightarrow \infty$.
To compensate their difference at the low-temperature region, we take $\Lambda_T= 0.130$ GeV for finite chemical potential.
As seen from Eq.(\ref{alphas0}), the effect owing to the different choice of $\Lambda_T$ in the high-temperature region is negligible.
Again, extrapolation is employed for temperature $T\lesssim \Lambda_T$.

\section{The hadronic resonance gas model}

The pressure of HRG with excluded volume correction~\cite{Rischke:1991ke} can be dertermined by the following self-consistent equations
\begin{eqnarray}
p^{H}(T,\mu_B,\mu_S,\mu_3)= \sum_{i=1} p_{i}^{id}(T,\widetilde{\mu_{i}}) ,\label{hrgeq}\\
\widetilde{\mu_{i}} \equiv  \mu_i - v_i p^{H} .\nonumber
\end{eqnarray}
In~\cite{Hama:2004rr}, the excluded volume $v_{i}=(4 \pi r_{0}^{3}/3)$, with $r_{0}=0.7fm$ for baryons and $r_{0}=0$ for mesons.

In the case of zero baryonic and strangeness density, one has $\mu_B=\mu_S=0$.
However, at finite baryon density, even though the strangeness density is zero, the strangeness chemical potential does not necessarily vanish.
This is because in this case the net strangeness density from baryons and their anti-patticles does not vanish at zero strangeness chemical potential, namely, the net strangeness density $n_S(\mu_B(\ne 0),\mu_S=0) - n_{S}(\mu_B\rightarrow \mu_B,\mu_S=0) \ne 0$.
Thus the value of strangeness chemical potential has to be determined by solving Eq.(\ref{hrgeq}) numerically.

We note that some improved HRG model with excluded volume correction has been proposed recently.
For instance, in Ref.~\cite{Vovchenko:2016rkn}, the authors considered not only the repulsive part of van der Waals interaction, but also the attractive part.
They found that the inclusion of van der Waals interaction leads to important implications for second and higher moments of fluctuations of conserved charges, in particular in the crossover region.
As in our model, the properties of the transition region is mostly determined by the lattice data, and there is no significant deviation between the models in the low-temperature region, the HRG model used in~\cite{Hama:2004rr} is adopted for our present study.

\section{Transition region and the implementation of the phenomenological critical point}

If the phase transition is of the first order, the chemical potential and temperature of the two phases are determined by the Gibbs condition.
In order to describe a smooth crossover in the region of small baryon density, we adopt the following scheme~\cite{Hama:2005dz}
\begin{eqnarray} \label{Gibbs}
(p-p^Q)(p-p^H)=\delta(\mu,T) ,
\end{eqnarray}
where
\begin{eqnarray}
\delta(\mu,T)=\delta_{0}(T) \exp \left[-(\mu/\mu_c)^4\right] ,
\end{eqnarray}
and $\mu_c$ is the critical chemical potential, which is taken to be $\mu_c = 0.3$ GeV in this work.

Eq.(\ref{Gibbs}) can be solved straightforwardly and one finds,
\begin{eqnarray}
p=\lambda p^H + (1-\lambda) p^Q+ \frac{2 \delta}{\sqrt{(p^Q-p^H)^2 + 4 \delta}} ,
\end{eqnarray}
where,
\begin{eqnarray}
\lambda = \frac{1}{2} [1-\frac{p^Q-p^H}{\sqrt{(p^Q-p^H)^2 + 4 \delta}}] \label{lambdasol}.
\end{eqnarray}
Other thermodynamic quantities can be obtained in terms of the grand partition function $q=\ln Q_G =\frac{pV}{k_{B}T}$.
Subsequently, one finds
\begin{eqnarray}
&&s=\lambda s^H + (1-\lambda) s^Q,\\
&&n_B=\lambda n^{H}_{B} + (1-\lambda) n^{Q}_{B} - \frac{2 \delta (\mu/\mu_c)^2}{\sqrt{(p^Q-p^H)^2 + 4 \delta}},\\
&&\varepsilon=\lambda \varepsilon^H + (1-\lambda) \varepsilon^Q - \frac{2 \delta (1+(\mu/\mu_c)^2)}{\sqrt{(p^Q-p^H)^2 + 4\delta}} .
\end{eqnarray}

We note, when $\delta_0=0$, a first order phase transition is recovered.
To be specific, we have $\lambda=0, p=p^Q$ when $p^Q > p^H$ and $\lambda=1, p=p^H$ when $p^Q < p^H$.
On the other hand, when $\delta(\mu_b) \ne 0$, the phase transition is smoothed out by an interpolation between the two phases.
In other words, instead of a sudden jump, $\lambda$ varies continuously from $0$ to $1$ during the transition.
Also, it is readily to verify that the above expression guarantees that the resulting pressure satisfies $p > p^Q$ and $p > p^H$.
Though strangeness chemical potential is considered in the model, we only consider the case of strange neutrality, and therefore, strangeness chemical potential is not a free parameter.

According to Eq.(\ref{lambdasol}), the ratio of $\delta$ to $|p^Q-p^H|$ determines whether the relevant thermodynamic quantities is dominated by one phase or more of a mixture of two phases.
We note that the Gibbs condition implies $|p^Q-p^H|=0$, and consequently, $|p^Q-p^H|$ becomes non-zero and increases once the system evolves away from the two-phase equilibrium.
In particular, $p^Q-p^H$ possesses different signs on different sides of the transition point.
As a result, the interpolation should work without any intervention as it is intended.
However, in practice, it is found that sometimes the magnitude of $|p^Q-p^H|$ decreases again as the system moves further away from the line of the first order phase transition, which may potentially jeopardize the interpolation scheme.
In a view to amending this issue, one defines a temperature interval so that the size of $\delta$ is suppressed on the outside of this region.
This is achieved by choosing $\delta_{0}(T)$ to be a piecewise function as follows
\begin{itemize}
  \item $\delta_{0}(T)=\delta_{0} e^{-c (T-T_{p})^{2}}, ~~~~~~~~T \leq T_{p}$
  \item $\delta_{0}(T)=\delta_{0},~~~~~~~~~~~~~~~~~~~~T_{p}<T \leq T_{p}+0.02$
  \item $\delta_{0}(T)=\delta_{0} e^{-c (T-T_{p}-0.02)^{2}},~~T > T_{p}+0.02$
\end{itemize}
where $\delta_{0}=5.90 \times 10^{-10}$ GeV$^8$ and $c=10^{3}$.
$T_{p}$ stands for the temperature (in GeV) of the corresponding first order transition.

\section{Numerical Results}

In this section, we present the numerical results of the obtained EoS by using the parameters summarized in Table \ref{coef}.

\begin{figure}[htb]
\begin{center}
\includegraphics[width=250pt]{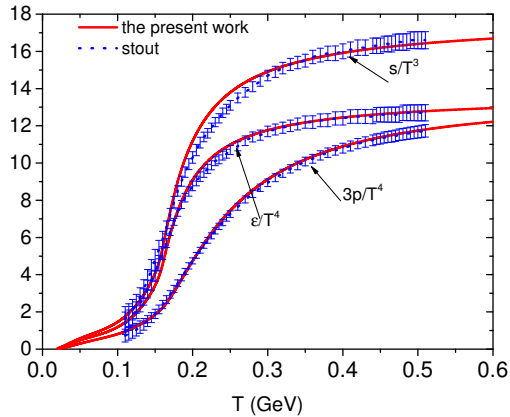} \\
\end{center}
\caption{(Color online) The calculated of $3p/T^4$ , $\varepsilon/T^4$ and $s/T^3$ using the quasiparticle model in comparison with those by lattice QCD with stout action~\cite{Borsanyi:2012cr,Borsanyi:2013bia} at zero chemical potential.}
\label{pes}
\end{figure}

\begin{table}
\begin{center}
\caption{List of parameters used in the present hybrid EoS} \vspace{0.5cm}
\begin{tabular}{|c|c|c|c|c|c|}
   \hline
    $T_{0}$ (GeV) & $p_{0}/T_{0}^{4}$ & $\Lambda_{T}$ for $\mu_B=0$ (GeV)  & $\Lambda_{T}$ for $\mu_B\ne 0$ (GeV)  & $\mu_{c}$ (GeV) & $c$ (GeV$^{-2}$)  \\
   \hline
   0.175 & 1.08    & 0.135  & 0.130  & 0.3  & $10^{3}$ \\
  \hline
   $\delta_{0}$ (GeV$^8$) & $m_{s0}$ (GeV) & $m_{u0,d0}$ (GeV) & $a_{g}^{2}$  & $a_{q}^{2}$  & $b_{q}^{2}$ \\
  \hline
   $5.90 \times 10^{-10}$ &0.15 & $5.33 \times 10^{-4}$  & 0.171 & 0.101  & 0.304 \\
  \hline
\end{tabular}
\label{coef}
\end{center}
\end{table}
As in \cite{Schneider:2001nf,Ivanov:2004gq}, an overall normalization factor 1.06 is introduced to take into account the unknown correction to the effective number of degrees of freedom.
For zero chemical potential, the resulting entropy density, energy density, and pressure are shown in Fig.\ref{pes} in comparison with the lattice QCD results in dotted blue curves with uncertainties~\cite{Borsanyi:2012cr,Borsanyi:2013bia}.
We see that all three quantities are reasonably well reproduced.

\begin{figure}
\begin{tabular}{cc}
\begin{minipage}{250pt}
\centerline{\includegraphics[width=250pt]{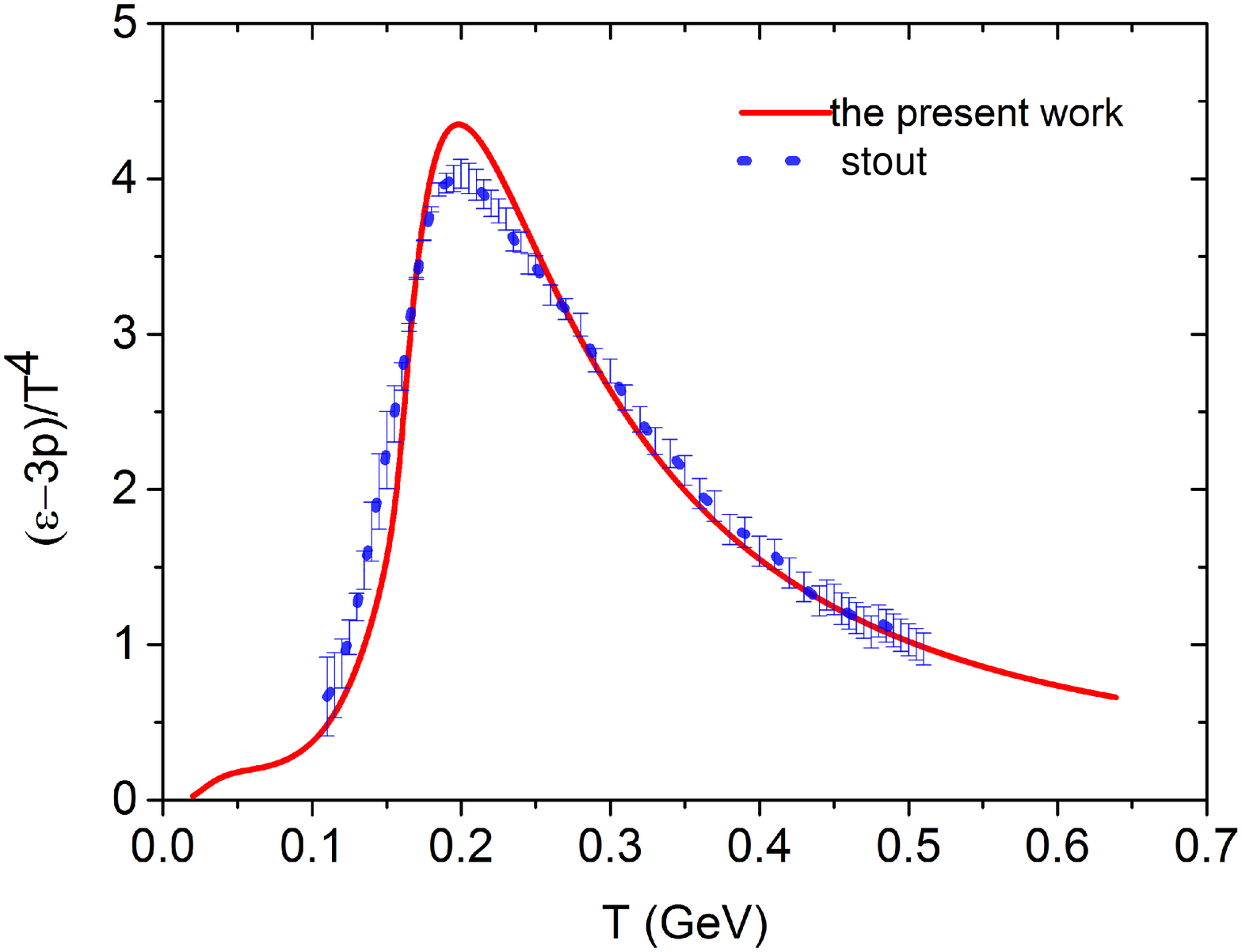}}
\end{minipage}
&
\begin{minipage}{250pt}
\centerline{\includegraphics[width=250pt]{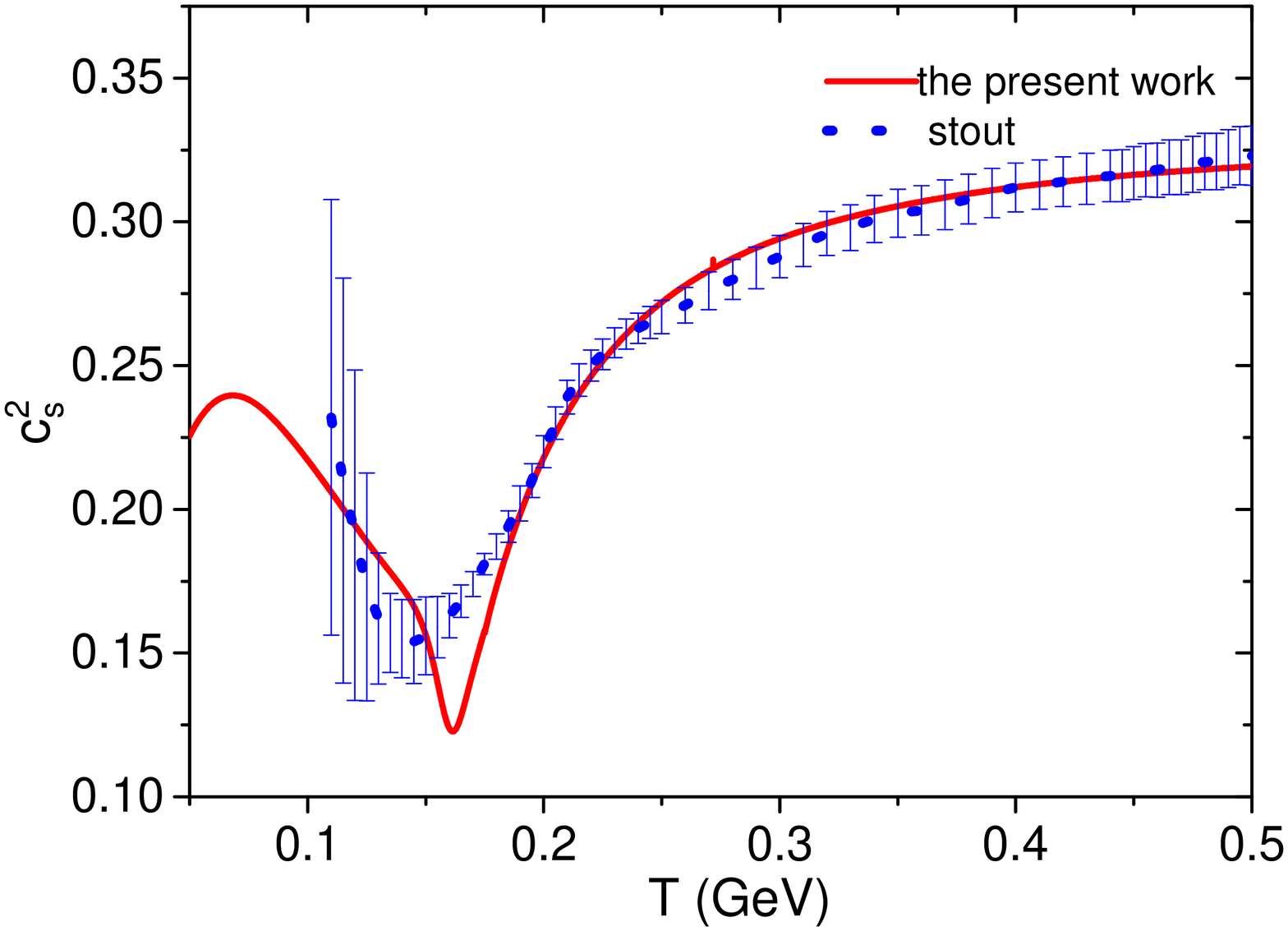}}
\end{minipage}
\end{tabular}
 \caption{(Color online) The calculated results in comparison with the lattice QCD data~\cite{Borsanyi:2012cr,Borsanyi:2013bia}.
 (a) the trace anomaly as a function of temperature,  (b) the speed of sound as a function of temperature.}
 \label{tracecs2}
\end{figure}

Another physical quantity of interest is trace anomaly, which is a measure of deviation from the conformal symmetry.
By lattice QCD simulations, the square of the speed of sound, $c_s^2=\frac{\partial p}{\partial \epsilon}$, is found to be smaller than that of an ideal gas of massless particles.
In particular, it is found that as $T$ approaches the transition region, $c_{s}^{2}$ reaches down to a minimum and then increases again in accordance with the HRG description of the system.
Since the above properties have potentially observable consequences during the hydrodynamical expansion of the system, it is, therefore, an important feature for the EoS.
The calculated trace anomaly and the sound are presented in Fig.\ref{tracecs2}.
It is found that the trace anomaly is reasonably reproduced.
The main feature of the speed of sound is also obtained, though the location of the minimum is slightly shifted towards higher temperature.
We note that Fig.\ref{tracecs2} is completely determined by those presented in Fig.\ref{pes}.
In the case of trace anomaly, the maximum of the curve is near $T\sim 0.2$ GeV.
In this region, as seen in Fig.\ref{pes}, the present model reproduces the pressure well in this region but slightly overestimates the energy density. 
Moreover, the deviation of $\varepsilon$ from the lattice data increases with increasing temperature in the vicinity of $T\sim 0.2$ GeV.
As a result, the maximum of the calculated trace anomaly overestimate the lattice data and is slightly shifted towards the right.
On the other hand, since the speed of sound is related to the ratio of the derivatives of two curves in Fig.\ref{pes}, it is more sensitive to the specific parameterization.
To be specific, in the region $T\sim 0.15$ GeV, the derivative $d\varepsilon/dT$ slightly underestimates the data at low temperature, namely, the calculated curve $\varepsilon/T^4$ is a bit too flat comparing to the data and then it becomes steeper as the temperature increases, while $dp/dT$ behaves oppositely in this region.
Consequently, the calculated sound speed underestimates the lattice data and the minimum is slightly shifted to the right.
Since the properties of the system at $T\sim 0.15$ GeV is mostly determined by the HRG model, one observes that the use of a fine-tuned model might further improve the result.

\begin{figure}[htb]
\begin{center}
\includegraphics[width=250pt]{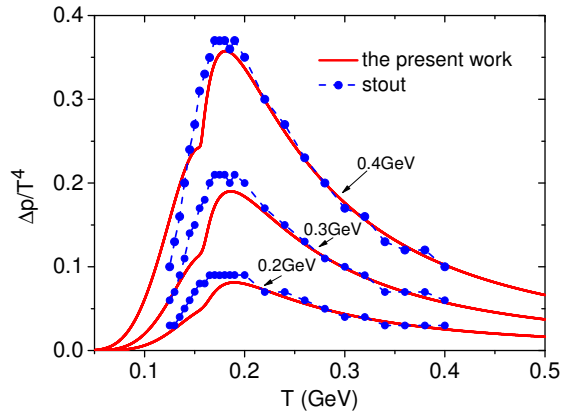} \\
\end{center}
\caption{(Color online) The calculated $\Delta p/T^4$ as a function of temperature for different chemical potentials, in comparision with the lattice QCD results by stout action~\cite{Borsanyi:2012cr}.}
\label{deltap}
\end{figure}

For finite chemical potential, pressure differences are calculated for different chemical potentials.
The results are shown in Fig.\ref{deltap}, in comparison with the lattice QCD results by stout action~\cite{Borsanyi:2012cr}.
As discussed above, our choice of $\Lambda_T$ ensures appropriate behavior at low temperature, while the results are insensitive to the specific value of $\Lambda_T$ at high temperature, due to Eq.(\ref{alphas0}).
The calculated results of $\Delta p$ agree well with the lattice data.

\begin{figure}
\begin{tabular}{ccc}
\begin{minipage}{160pt}
\centerline{{\includegraphics[width=160pt]{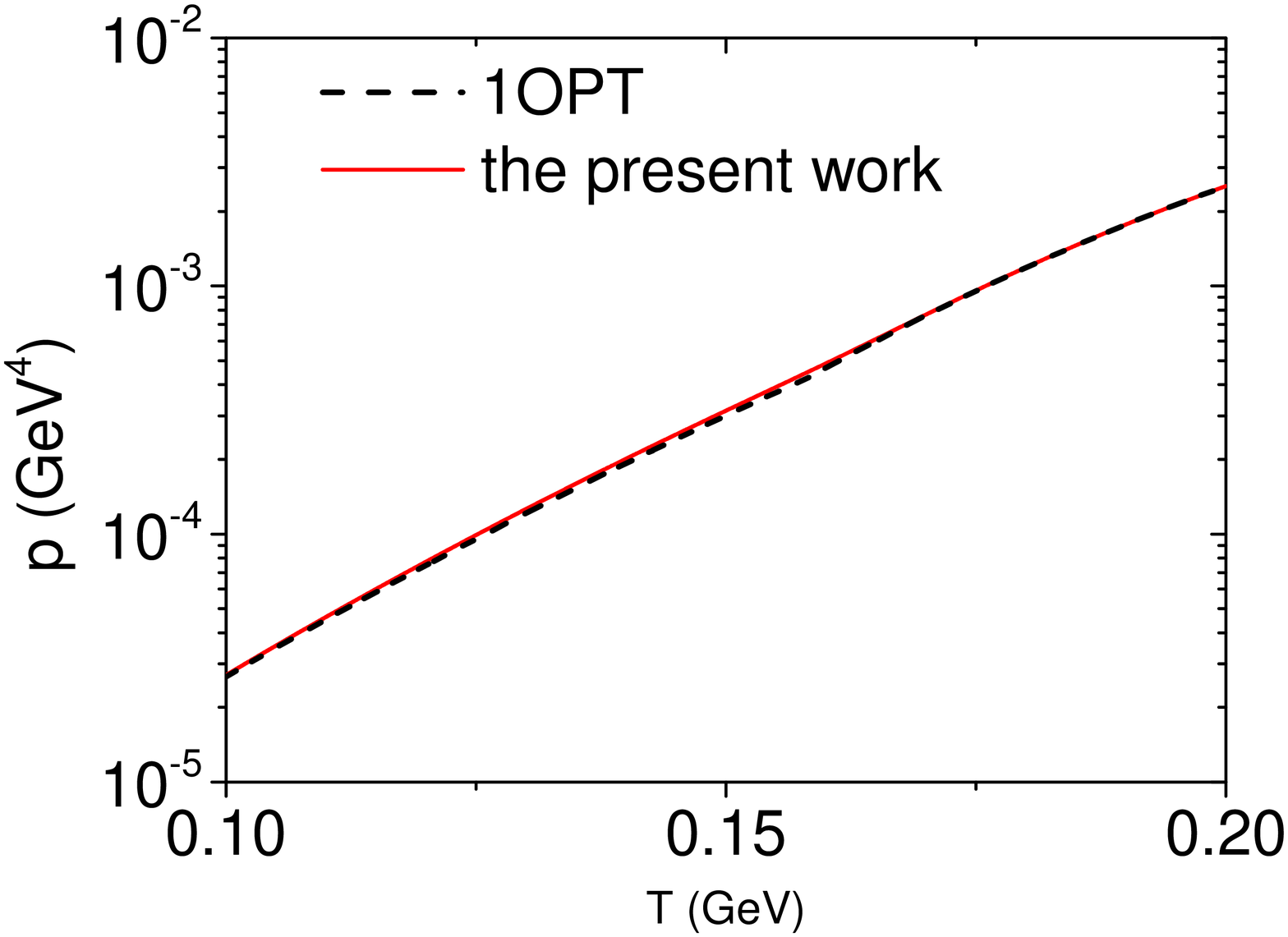}}}
\end{minipage}
&
\begin{minipage}{160pt}
\centerline{{\includegraphics[width=160pt]{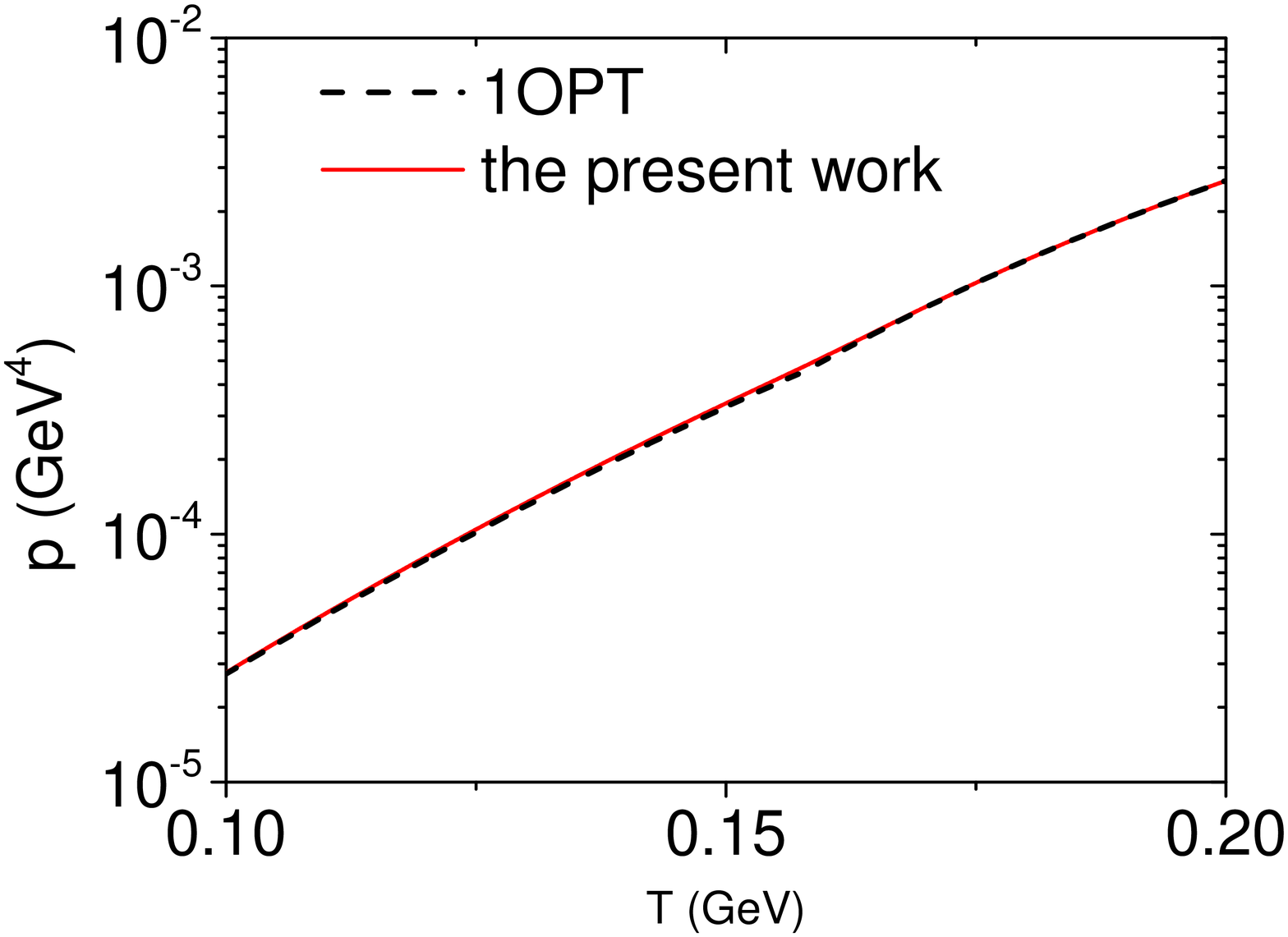}}}
\end{minipage}
&
\begin{minipage}{160pt}
\centerline{{\includegraphics[width=160pt]{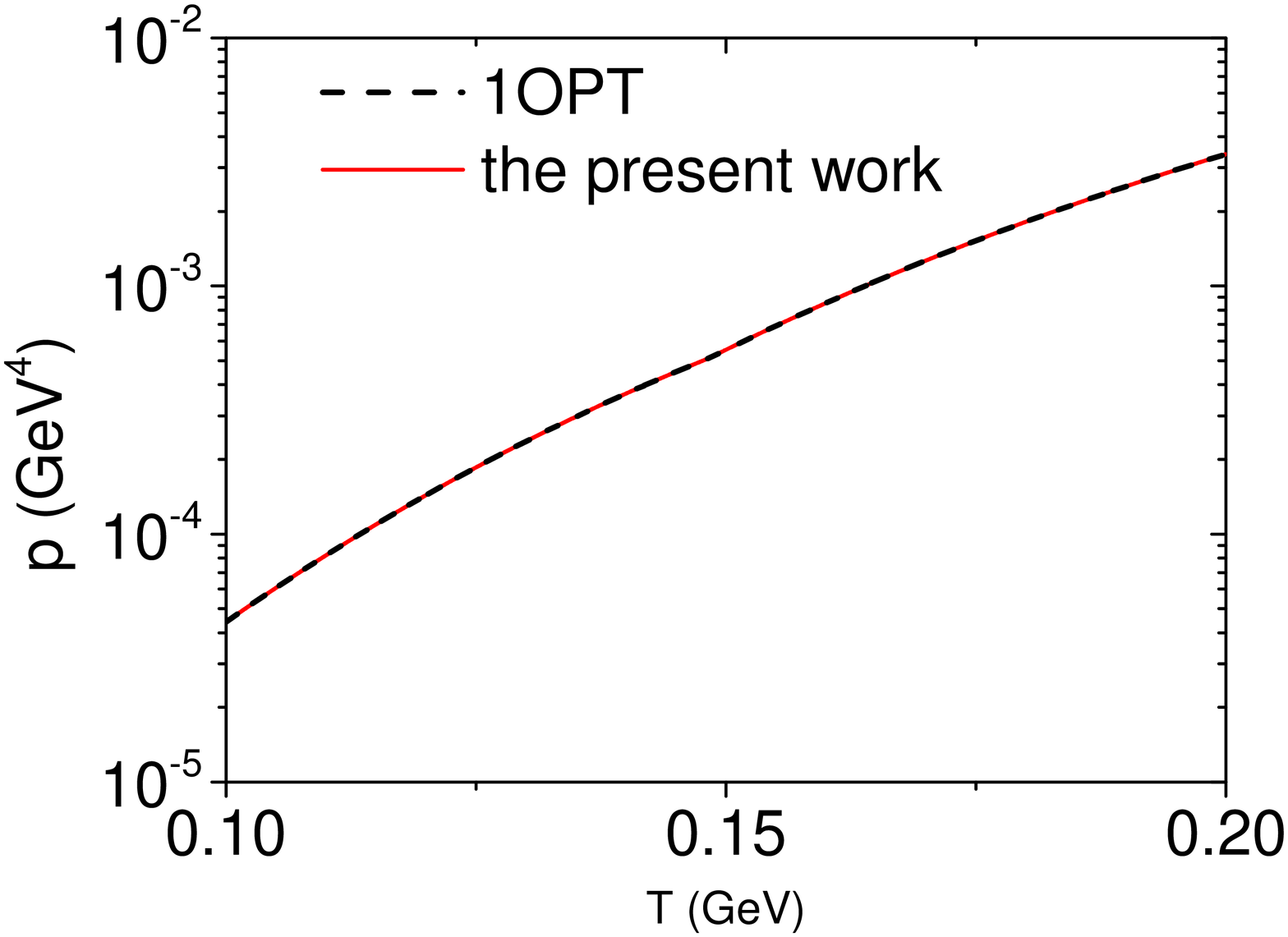}}}
\end{minipage}
\\
\begin{minipage}{160pt}
\centerline{{\includegraphics[width=160pt]{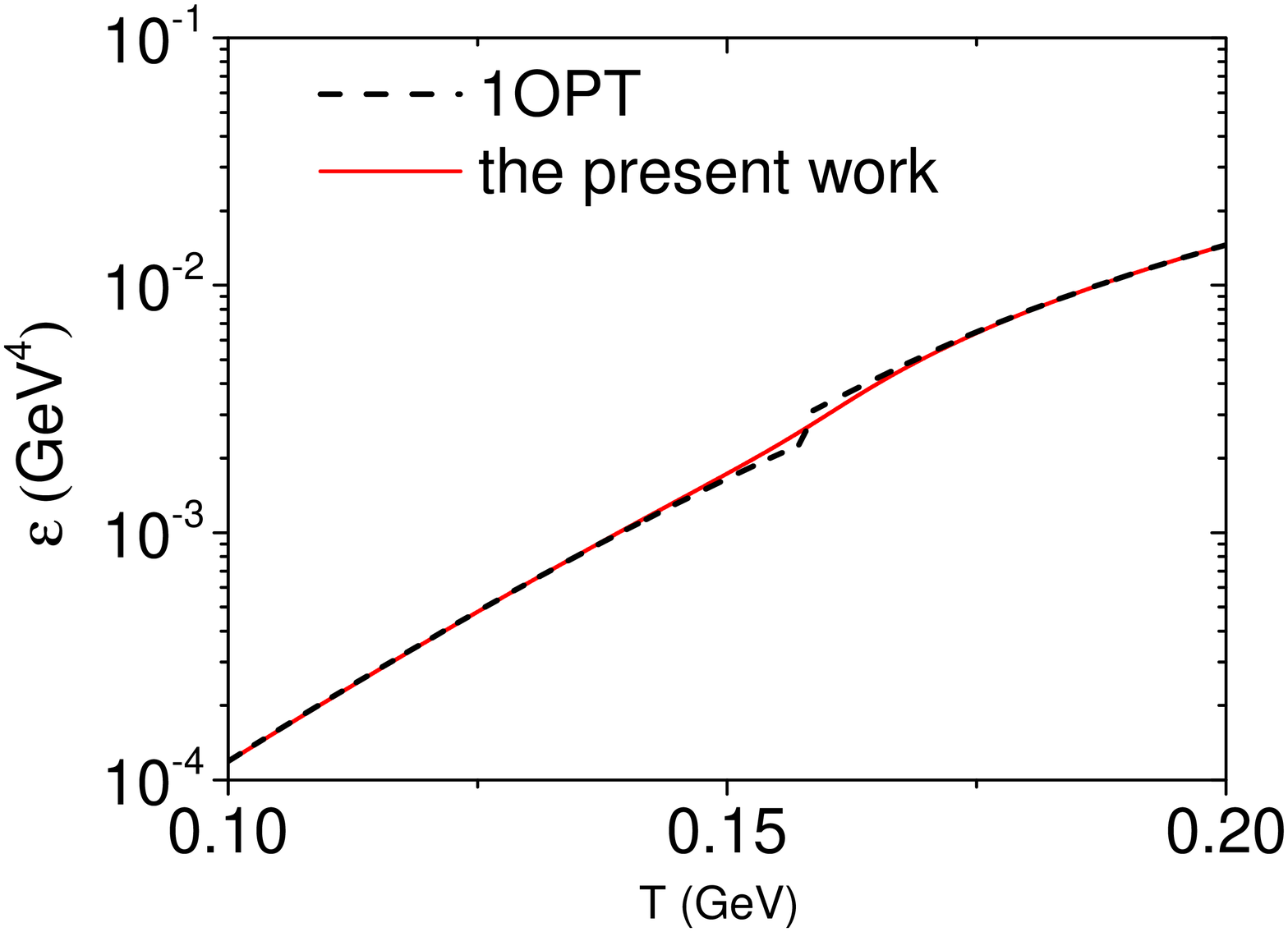}}}
\end{minipage}
&
\begin{minipage}{160pt}
\centerline{{\includegraphics[width=160pt]{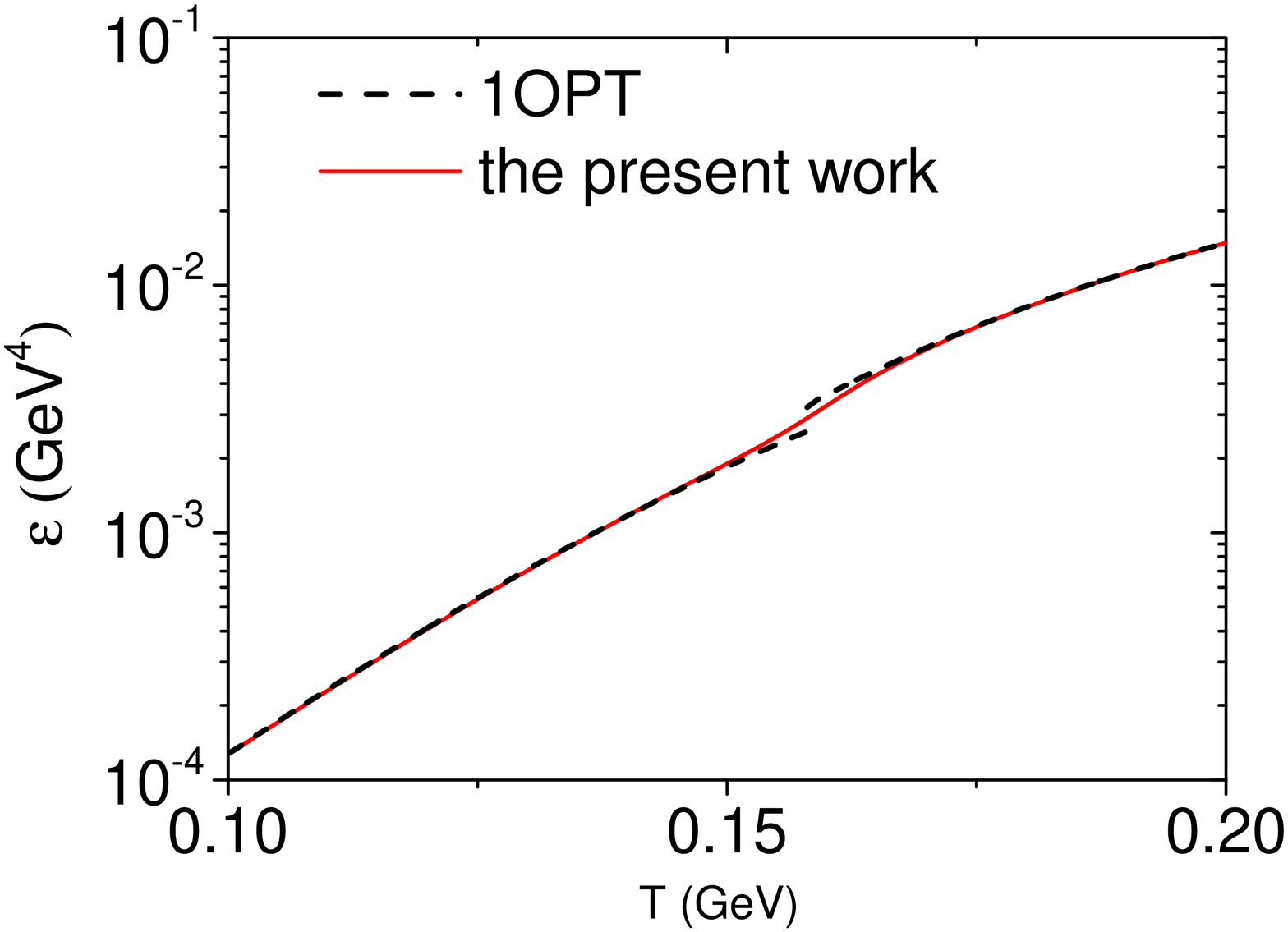}}}
\end{minipage}
&
\begin{minipage}{160pt}
\centerline{{\includegraphics[width=160pt]{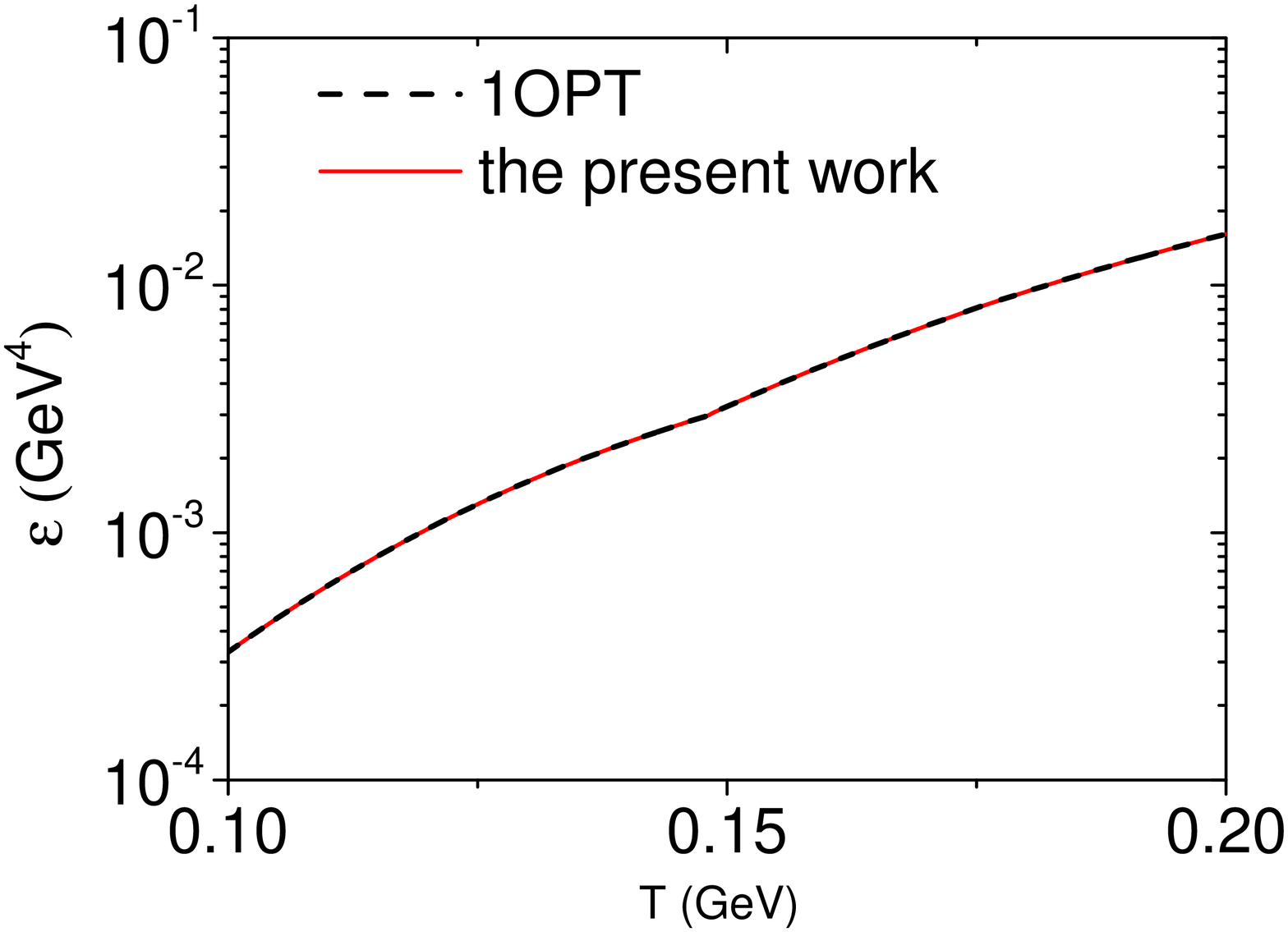}}}
\end{minipage}
\\
\begin{minipage}{160pt}
\centerline{{\includegraphics[width=160pt]{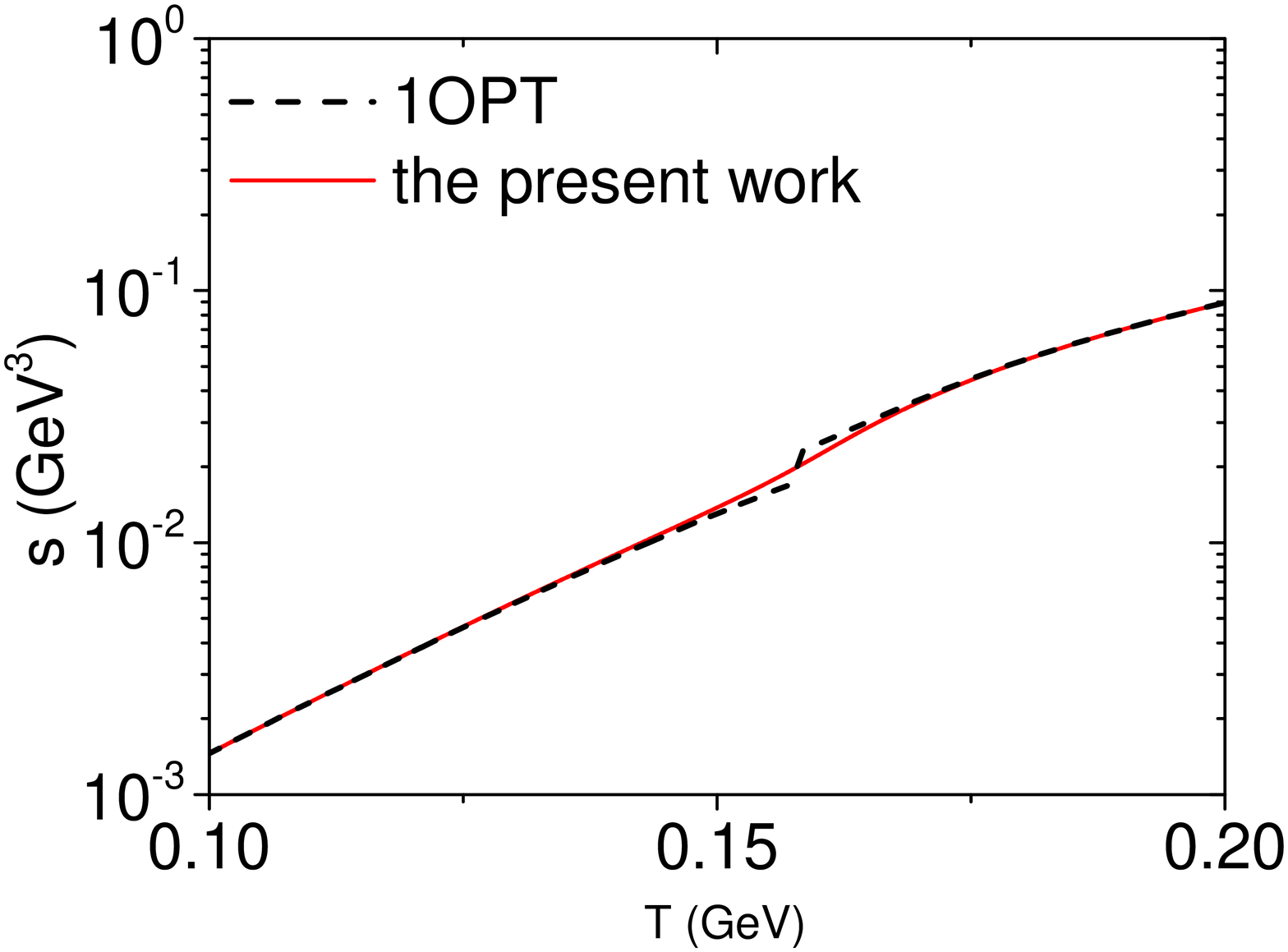}}}
\end{minipage}
&
\begin{minipage}{160pt}
\centerline{{\includegraphics[width=160pt]{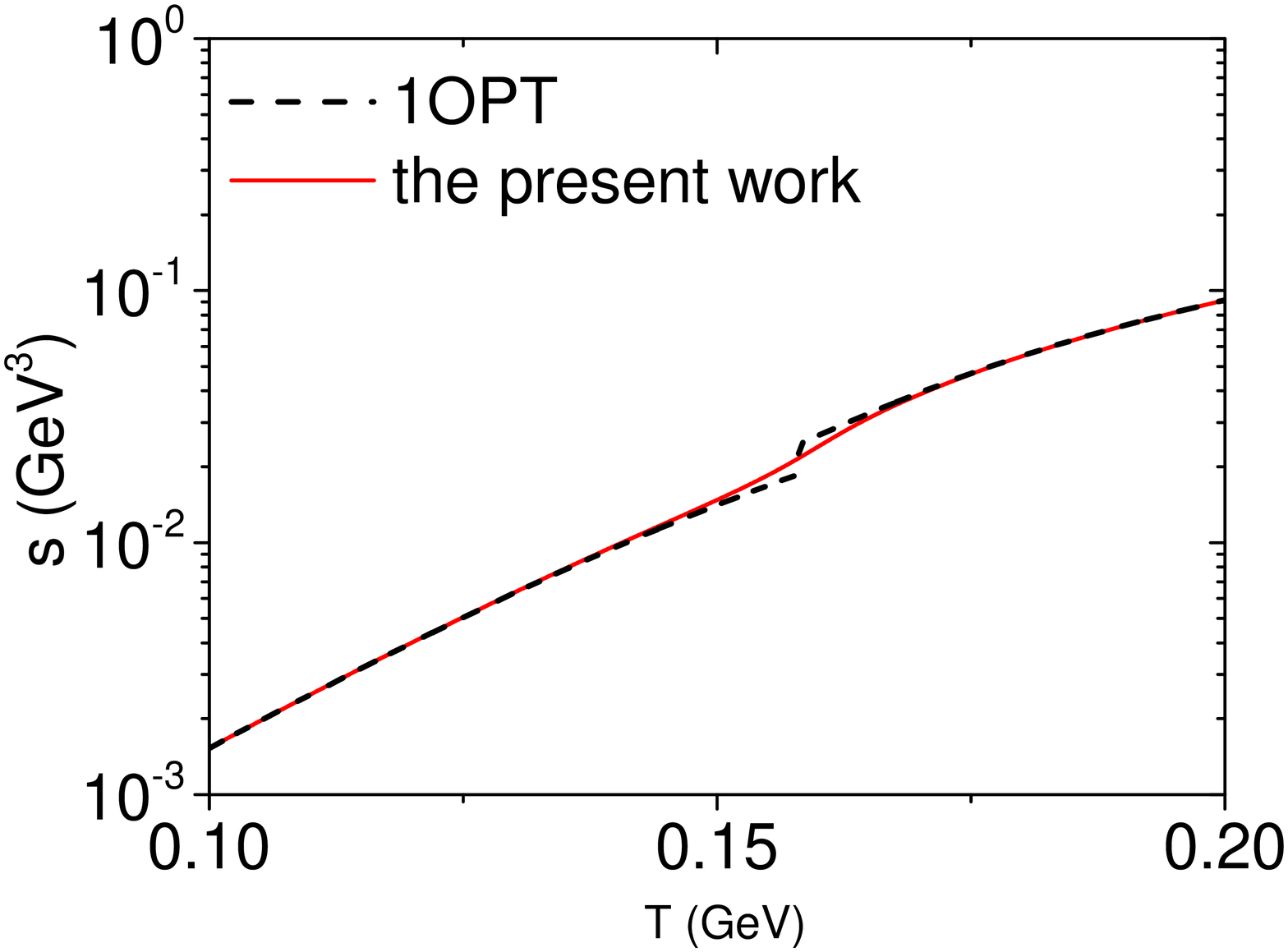}}}
\end{minipage}
&
\begin{minipage}{160pt}
\centerline{{\includegraphics[width=160pt]{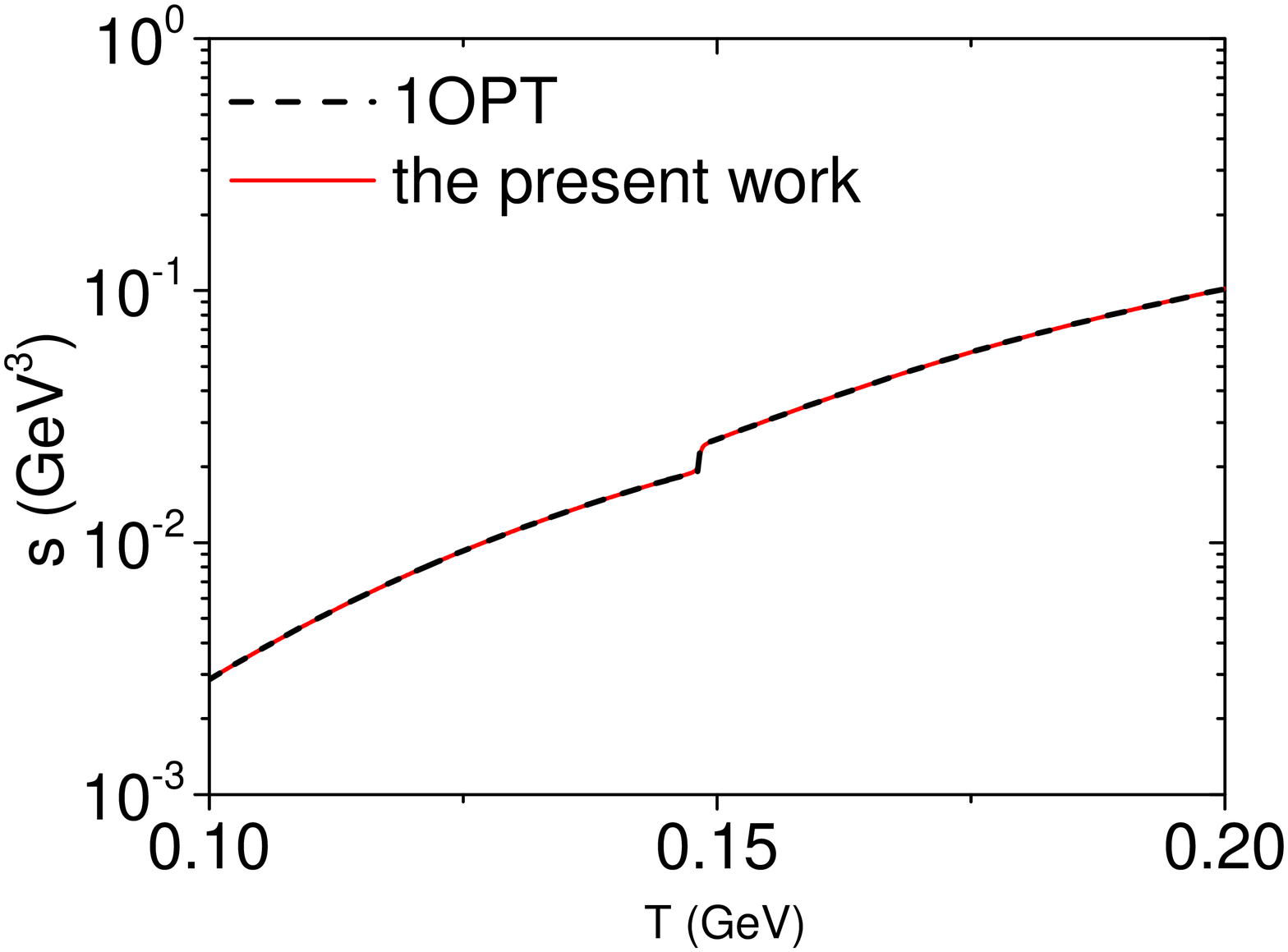}}}
\end{minipage}
\\
\begin{minipage}{160pt}
\centerline{\subfigure[$0.0$GeV]{\includegraphics[width=160pt]{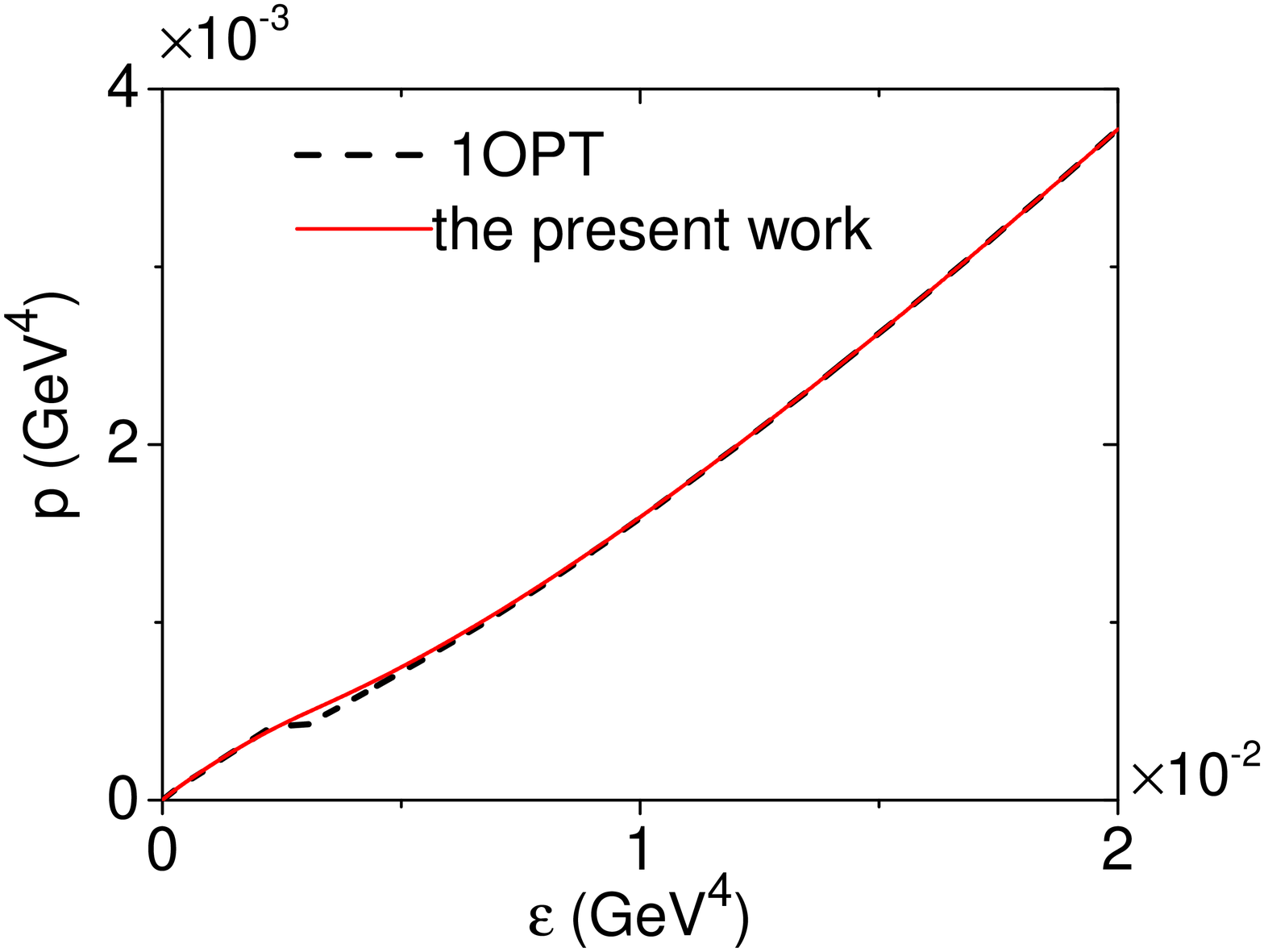}}}
\end{minipage}
&
\begin{minipage}{160pt}
\centerline{\subfigure[$0.2$GeV]{\includegraphics[width=160pt]{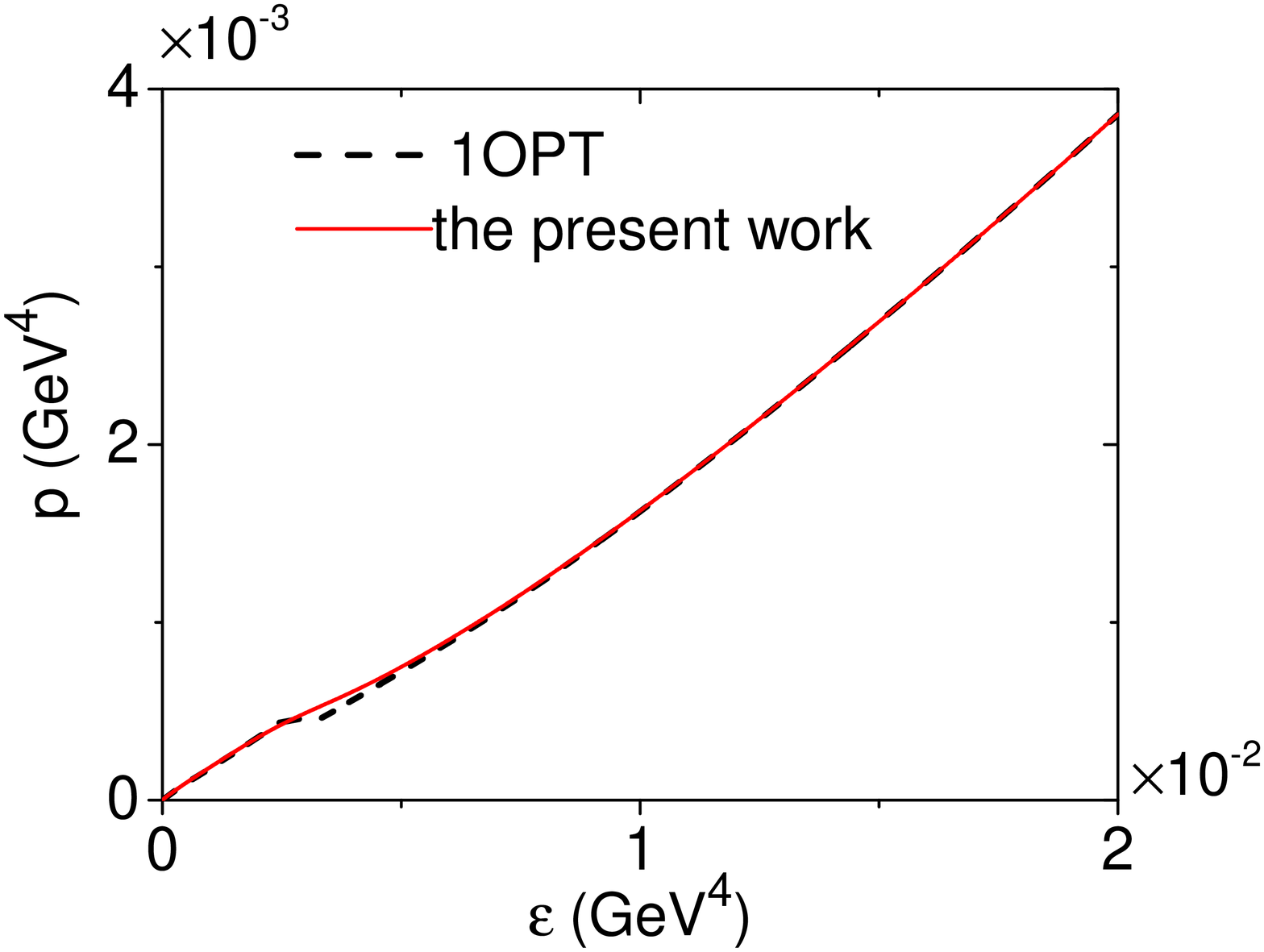}}}
\end{minipage}
&
\begin{minipage}{160pt}
\centerline{\subfigure[$0.5$GeV]{\includegraphics[width=160pt]{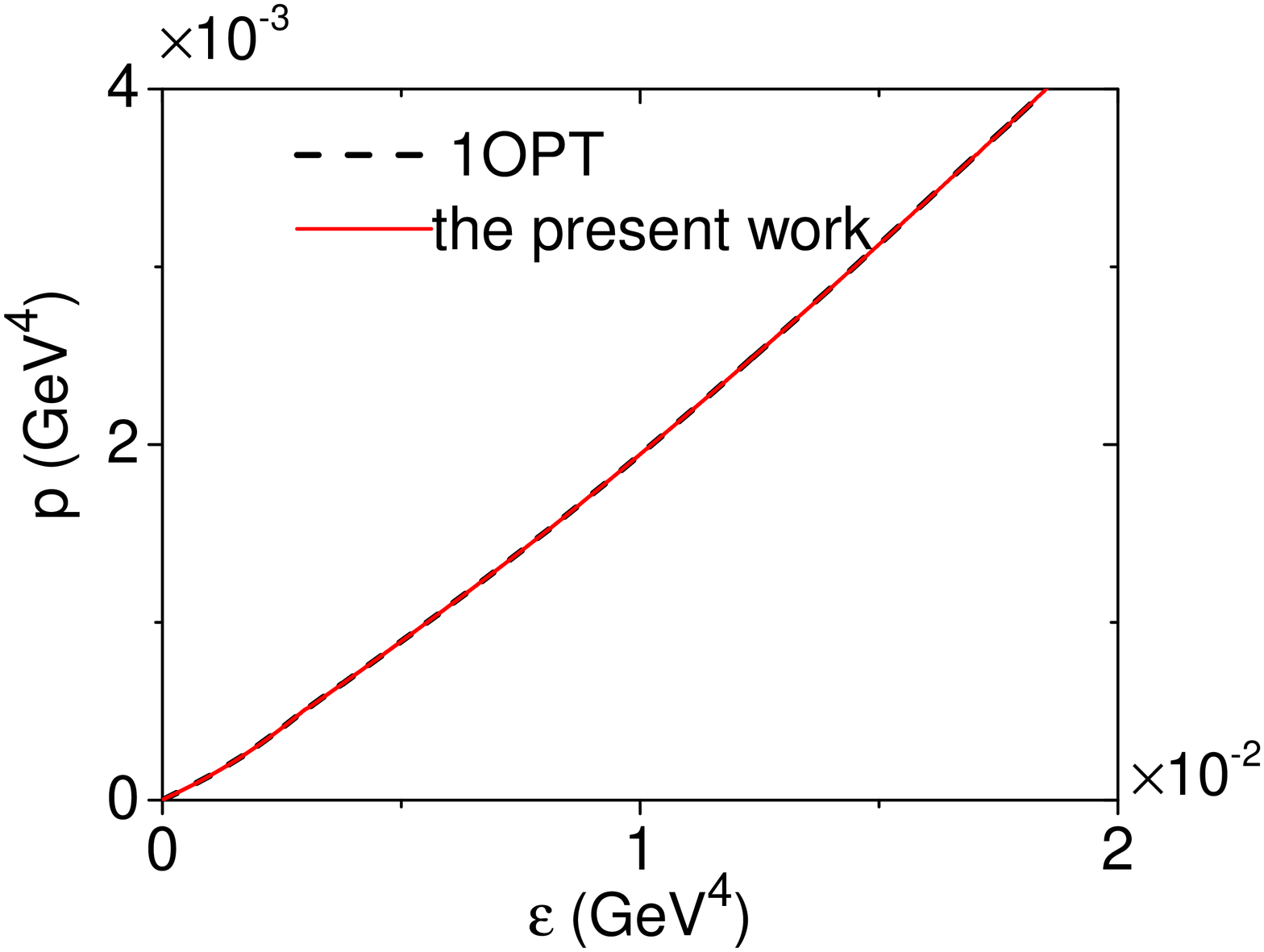}}}
\end{minipage}
\\
\end{tabular}
\caption{(Color online) The calculated the pressure, energy density and entropy density as functions of temperature, and pressure as function of energy density for different chemical potentials.
The results of the present interpolation scheme is compared to those of a first order phase transition (1OPT). From left to right, the three columns are for $\mu_B=0, 0.2 $ and $0.5$ GeV respectively.}
\label{eosvs}
\end{figure}

Now, we compare the results of the proposed model with those obtained by first-order phase transition.
For $\mu_B < \mu_c$, the present interpolation scheme gives a smooth crossover in the transition region, which is distinct from that of a first order phase transition.
When one goes beyond the critical chemical potential, the transition gradually approaches of that of a first order which involves discontinuous changes of thermodynamic quantities related to the first order derivative of the Gibbs thermodynamical potential.
At high temperature, the quasiparticle model guarantees that the results approach those of lattice QCD calculations.
This is shown in Fig.\ref{eosvs}, where we present the pressure, energy density and entropy density as functions of temperature, and pressure as functions of energy density for different chemical potentials.
It is found that for $\mu_B=0$, all physical quantities vary smoothly during the transition for the present interpolation scheme, while the results for the first order phase transition EoS show sudden jumps in energy density and entropy density.
Similar results are obtained for $\mu_B = 0.2 $ GeV, which is also below the critical chemical potential $\mu_c$.
On the other hand, for $\mu_B=0.5$ GeV, the obtained results are almost identical to those of the first order phase transition, as expected.

\section{Concluding remarks}

In this paper, an interpolation scheme is adopted to build an EoS with a phenomenological critical point at finite chemical potential.
A quasiparticle model is fitted to the lattice QCD data to describe the high-temperature QGP phase, while an HRG model with exclusive volume correction is utilized for the hadronic phase in the low-temperature region.
The critical point is implemented so that all other quantities are derived from the Gibbs thermodynamic potential, and therefore, the thermodynamic consistency is guaranteed in the present approach.

The EoS plays an essential role in the hydrodynamic description of relativistic heavy-ion collisions~\cite{sph-review-1}.
It directly affects many physical quantities which include particle spectrum~\cite{sph-eos-2,sph-cfo-1}, collective flow, and two-pion interferometry~\cite{sph-eos-3} among others.
In particular, the ongoing RHIC beam energy scan program is aimed to study the QCD phase boundary and search for the possible QCD critical point.
Obviously, the existence of a critical point will affect the temporal system evolution and subsequently various observables~\cite{star-bes-01,star-bes-02,star-bes-03,star-bes-04}, such as particle ration, multiplicity, as well as $p_T$ fluctuations, harmonic flow coefficients, and dihadron correlation.
We plan to carry out a hydrodynamic study of the relevant quantities using the present EoS shortly.

\section*{Acknowledgments}
We are thankful for valuable discussions with Yogiro Hama, Takeshi Kodama, and Pasi Huovinen.
We gratefully acknowledge the financial support from
Funda\c{c}\~ao de Amparo \`a Pesquisa do Estado de S\~ao Paulo (FAPESP),
Funda\c{c}\~ao de Amparo \`a Pesquisa do Estado do Rio de Janeiro (FAPERJ),
Conselho Nacional de Desenvolvimento Cient\'{\i}fico e Tecnol\'ogico (CNPq),
and Coordena\c{c}\~ao de Aperfei\c{c}oamento de Pessoal de N\'ivel Superior (CAPES).
This research is also supported by the Center for Scientific Computing (NCC/GridUNESP) of the S\~ao Paulo State University (UNESP).

\bibliographystyle{h-physrev}
\bibliography{references_ma,references_qian}

\end{document}